# Exploring Solute Behavior and Texture Selection in Magnesium Alloys at the Atomistic Level

F. Mouhib[1*], Z. Xie[1*], A. Atila[2], J. Guénolé[3], S. Korte-Kerzel[1], T. Al-Samman[1]

1. Institute of Physical Metallurgy and Materials Physics, RWTH Aachen University, 52056 Aachen, Germany

2. Department of Materials Science and Engineering, Saarland University, 66123 Saarbrücken, Germany

3. Université de Lorraine, CNRS, Arts et Métiers ParisTech, LEM3, 57070 Metz, France

[*]Corresponding authors, email mouhib@imm.rwth-aachen.de, xie@imm.rwth-aachen.de

**Abstract:**

This study advances our understanding of how chemical binding and solute distribution impact grain boundary segregation behavior and subsequent annealing texture modification in lean Mg-X-Zn alloys (X = RE or Ca). Notably, differences in Ca and Gd solute behavior at grain boundaries were revealed, where Ca exhibited stronger binding to vacancy sites than Gd, resulting in elevated Ca segregation and an RD-TD-type texture. The introduction of Zn showed significant synergistic effects on solute clustering, with Gd-Zn pairs forming more favorably than Ca-Zn pairs, leading to a strong synergy between Zn and Gd. This promoted their co-segregation and high concentration at the grain boundary, generating a unique TD-spread texture. In contrast, weaker binding in Ca-Zn pairs did not affect Ca segregation but influenced Zn segregation, which underscores the importance of solute binding behavior in alloy design concepts. Additionally, the combined atomic-scale experiments and ab initio predictions provide strong evidence that selective texture development in Mg alloys is tied to heterogeneous solute-boundary interactions, where the sensitivity of the binding energy to volumetric strain affects solute segregation at grain boundaries, resulting in varying grain boundary mobilities and specific texture component growth. It also emphasizes that solute behavior in clustering and segregation is influenced not only by atomic size but also by chemical binding strength with vacancies or co-added Zn.

## 1 Introduction

The choice of alloying elements hinges upon the specific requirements and performance criteria of a given application, where solute additions have demonstrated a remarkable capability to significantly enhance and modify the material properties. As such, they have become indispensable to the design of cutting-edge materials. Fundamental research in this area has shown that interactions between solutes and defects play a pivotal role in shaping microstructure formation and influencing the mechanical response of materials [1-3]. This is particularly important in the context of lightweight materials like magnesium (Mg), where dilute alloying with specific solute elements extends the scope of their utility by improving their formability at low temperatures [4].

The addition of rare earth elements (RE) to Mg alloys has proven to be a highly effective strategy for altering the material's texture and enhancing its mechanical properties. Several related studies have consistently reported enhanced ductility, increased yield strength, mitigated anisotropic behavior, and softened annealing textures that respond favorably to an applied strain, compared to common Mg alloys [5-7]. It was established that the large size misfit of RE solutes within the Mg matrix causes solute segregation to grain boundaries (GB) influencing the GB energy and mobility. Combined addition of transition elements, such as zinc (Zn) into Mg-RE alloys induces a negative lattice misfit, giving rise to local solute clusters of Zn and RE atoms with important co-segregation effects on the properties of grain boundaries [8, 9]. Numerous studies have demonstrated that the combination of RE elements with Zn intensifies the rare earth effect on texture, resulting in a unique split of basal poles into the transverse direction (TD) of rolled alloy sheets [7, 10-13]. In that respect, the available literature points to a correlation between GB solute segregation and an orientation selection process occurring during recrystallization nucleation and growth [9, 14]. Proposed mechanisms behind this phenomenon include anisotropic solute drag effects and alterations of the grain boundary structure and properties by the segregated solutes [2, 15-19]. A recent atom probe tomography (APT) study on Mg-Gd-Zn alloys has unveiled that the synergistic effects of combined solute species on GB segregation and texture selection are contingent not only on the type of solute atoms but also on the relative atomic concentration ratio between them [20].

Given the limited availability and higher cost of RE elements, calcium (Ca) frequently emerges as the preferred alloying element in Mg alloys because it is more abundant and cost-effective. Owing to its relatively large atomic size, it also tends to segregate to grain boundaries, and consequently produce qualitative changes of the texture components [21, 22]. However, its co-segregation behavior and associated GB drag effect in ternary alloys still lack clarity. As shown here, the resulting annealing texture in an Mg-Zn-RE alloy diverges, to some extent, from that in a counterpart Mg-Zn-Ca alloy, suggesting dissimilar segregation behaviors. Given that the atomic radius of Ca is similar to that of RE elements, the hypothesized disparity in the mechanistic factors governing texture development implies additional considerations beyond the elastic energy effect due to solute size mismatch. Hence, the present study endeavors to establish a better understanding of how solute interactions within the microstructure correlate with the macroscopic behavior of the investigated materials. This is pursued through a comprehensive approach that combines scale-bridging experiments and ab initio calculations, designed to provide and connect insights into the solute behavior and texture selection, spanning multiple scales of observation.

## 2 Experiments and Methods

### *2.1. Sample preparation*

Binary and ternary Mg-Gd-(Zn) and Mg-Ca-(Zn) were investigated in this study. Their chemical composition, measured by ICP/OES is given in Table 1. The alloys were melted in an induction furnace under a protective Ar/$CO_2$ gas atmosphere and subsequently cast into a pre-heated copper mold to be finally homogenized at 460°C for 960 min.

Table 1. Chemical composition of the investigated binary and ternary alloys and their

| Alloys | Gd (at %) | Ca (at %) | Zn (at %) |
|---|---|---|---|
| Mg-Gd | 0.142 | - | - |
| Mg-Ca | - | 0.150 | - |
| Mg-Gd-Zn | 0.148 | - | 0.165 |
| Mg-Ca-Zn | - | 0.150 | 0.395 |

Sheet samples of the dimensions 60 × 40 × 4 $mm^3$ were machined out of the cast materials and hot rolled at 400°C (nominal furnace temperature) to 80% thickness reduction in multiple passes. To refine the microstructure, the final rolling pass was performed at a lower temperature (200°C). Smaller samples (10 x 12 x 1 $mm^3$) cut from the rolled sheets were annealed at 400°C for 60 min in a Heraeus RL200E air furnace. Metallographic sample preparation involved mechanical grinding and polishing of the mid-surface with a diamond suspension up to 0.25 μm. Additionally, all samples were electro-polished in a Struers AC-2 solution using a Lectro-Pol 5 operated at −20 °C and 25 V for 120 s.

### *2.2. Microstructure characterization*

For macrotexture characterization, six incomplete pole figures [$\{10\bar{1}0\}$, $\{0002\}$, $\{10\bar{1}1\}$, $\{10\bar{1}2\}$, $\{11\bar{1}0\}$, $\{10\bar{1}3\}$] were measured using a Bruker D8 advance diffractometer. From these incomplete pole figures, full pole figures and orientation distribution functions (ODFs) were calculated using the texture analysis toolbox MTEX [23]. A dual beam microscope Helios 600i equipped with a HKL-Nordlys II EBSD detector operating at 20 kV was used for EBSD measurements. The used step size varied between 0.5 and 1.5 μm depending on the grain size. The MTEX toolbox was used to analyze the raw EBSD data.

Site-specific preparation of APT tips with general, high-anle grain bounadries, identified through the EBSD measurements, was performed by a coordinated process of focused ion beam (FIB) milling and transmission Kikuchi diffraction (TKD) to ensure a precise position of the GB within the APT tips. The FIB sharpening of the tips was carried in several steps with decreasing beam currents (from 0.43 nA to 40 pA) and milling radii (from 1.5 to 0.1 nm). Subsequently, the elemental distributions of solute atoms in the vicinity of the measured grain boundaries were characterized by 3D APT using a Cameca LEAP 4000X HR in laser-pulsing mode at 30 K. This involved an ultra-violet laser with wavelength of 355 nm, pulse energy of 30 pJ and a pulse rate of 125 kHz maintaining a constant evaporation rate of 0.5%.

*2.3. Local solute cluster analysis*

To assess the local clustering tendency of multiple solute elements in the reconstructed APT samples, a clustering ratio parameter $R_{A-B}$ is utilized [24].This parameter compares the number of atoms of a particular solute element within a specific cutoff distance $r$ to the expected number of atoms of that element in a homogeneous distribution within the same cutoff. Accordingly, the average number of B atoms within a specific cutoff distance from an A site is calculated using the following equation:

$$\bar{N}_{A-B}(r) = \frac{\sum_{N_A} N_{A-B}(r)}{N_A},$$

where $N_{A-B}(r)$ is the number of B atoms surrounding an A site within $r$ (excluding a potential B atom on the A site) and $N_A$ is the number of A sites. Assuming a homogeneous distribution of B atoms throughout the sample, the ideal number of B atoms within a volume with a cutoff distance $r$ can be calculated as:

$$N_{B,homo}(r) = \frac{4}{3}\pi r^3 \frac{N_B}{V},$$

From that, the clustering ratio $R_{A-B}$ is defined as:

$$R_{A-B} = \frac{\bar{N}_{A-B}(r)}{N_{B,homo}(r)},$$

if $R_{A-B} = 1$, the distribution of B atoms surrounding the A site within a cutoff distance $r$ is considered homogeneous.

For the analysis, the reconstructed APT tips were divided into two regions: a GB region with a thickness of 20 nm along the GB, and a bulk region comprising the remaining part. To calculate the $R_{A-B}$ values for the GB and bulk regions, as well as for the entire tip, the data was normalized using $N_{B,homo,bulk}(r)$, which is obtained based on the number of B atoms in the bulk region, $N_{B,bulk}$, and the corresponding volume, $V_{bulk}$.

*2.3. Density functional theory (DFT) calculations*

DFT calculations in the present work were done with the Quantum ESPRESSO package [25] using the projector augmented wave (PAW) [26, 27] method and the Perdew-Burke-Ernzerhof (PBE) generalized gradient approximation (GGA) [28] to the exchange-correlation functional. The kinetic energy cutoff was 40 Ry for wave functions and 280 Ry for charge densities and potentials. The convergence threshold for electronic self-consistency was $10^{-8}$ Ry. For RE solutes, 4f electrons were treated as core electrons in the pseudopotentials. The Broyden-Fletcher-Goldfarb-Shanno (BFGS) relaxation scheme was used for structural relaxation [29]. The convergence thresholds of energy and force are $10^{-4}$ a.u and $2\times10^{-4}$ a.u, respectively. The orthorhombic simulation cells consisted of $4 \times 2 \times 2$ unit cells (64 atoms) with a $6 \times 6 \times 6$ *k*-point mesh and $5 \times 3 \times 3$ unit cells (180 atoms) with a $4 \times 4 \times 4$ *k*-point mesh. The solute-vacancy and solute-solute binding energies were calculated by the following equations:

$$E_{\mathrm{bind}}^{\mathrm{vac-X}} = \left(E_{\mathrm{bulk}}^{\mathrm{vac}} + E_{\mathrm{bulk}}^{\mathrm{X}}\right) - \left(E_{\mathrm{bulk}}^{\mathrm{vac-}} \quad + E_{\mathrm{bulk}}\right), \qquad \text{(Eq. 2.1)}$$

$$E_{\mathrm{bind}}^{\mathrm{X-Y}} = \left(E_{\mathrm{bulk}}^{\mathrm{X}} + E_{\mathrm{bulk}}^{\mathrm{Y}}\right) - \left(E_{\mathrm{bulk}}^{\mathrm{X-Y}} + E_{\mathrm{bulk}}\right), \qquad \text{(Eq. 2.2)}$$

where $E_{\mathrm{bulk}}$ is the energy of the Mg matrix, $E_{\mathrm{bulk}}^{\mathrm{vac,X,Y}}$ is the energy of the Mg matrix containing one vacancy or a solute atom X or Y, $E_{\mathrm{bulk}}^{\mathrm{vac-X}}$ is the energy of the Mg matrix containing a pair of vacancy and solute atom X, and $E_{\mathrm{bulk}}^{\mathrm{X-Y}}$ is the energy of the Mg matrix containing a pair of solute atoms X and Y. In the current calculations, a positive binding energy indicates attraction, i.e. a favorable binding. The energies were calculated at 0 K.

## 3 Results

### *3.1. Annealing textures*

The development of the annealing texture in the Gd-containing alloys reveals notable distinctions in comparison to the counterpart Ca-containing alloys. Fig. 1 shows the (0002) pole figures of the examined binary and ternary Mg-Gd-(Zn) and Mg-Ca-(Zn) alloys after 80% rolling and subsequent recrystallization annealing at 400°C for 60 min. In case of the Gd-containing alloy there is a pronounced texture transition from an RD-split texture to a TD-split texture upon the addition of Zn. However, this trend is less evident in the Mg-Ca and Mg-Ca-Zn alloys, where the spread of basal poles and associated maximum densities are almost equal in the RD and TD directions.

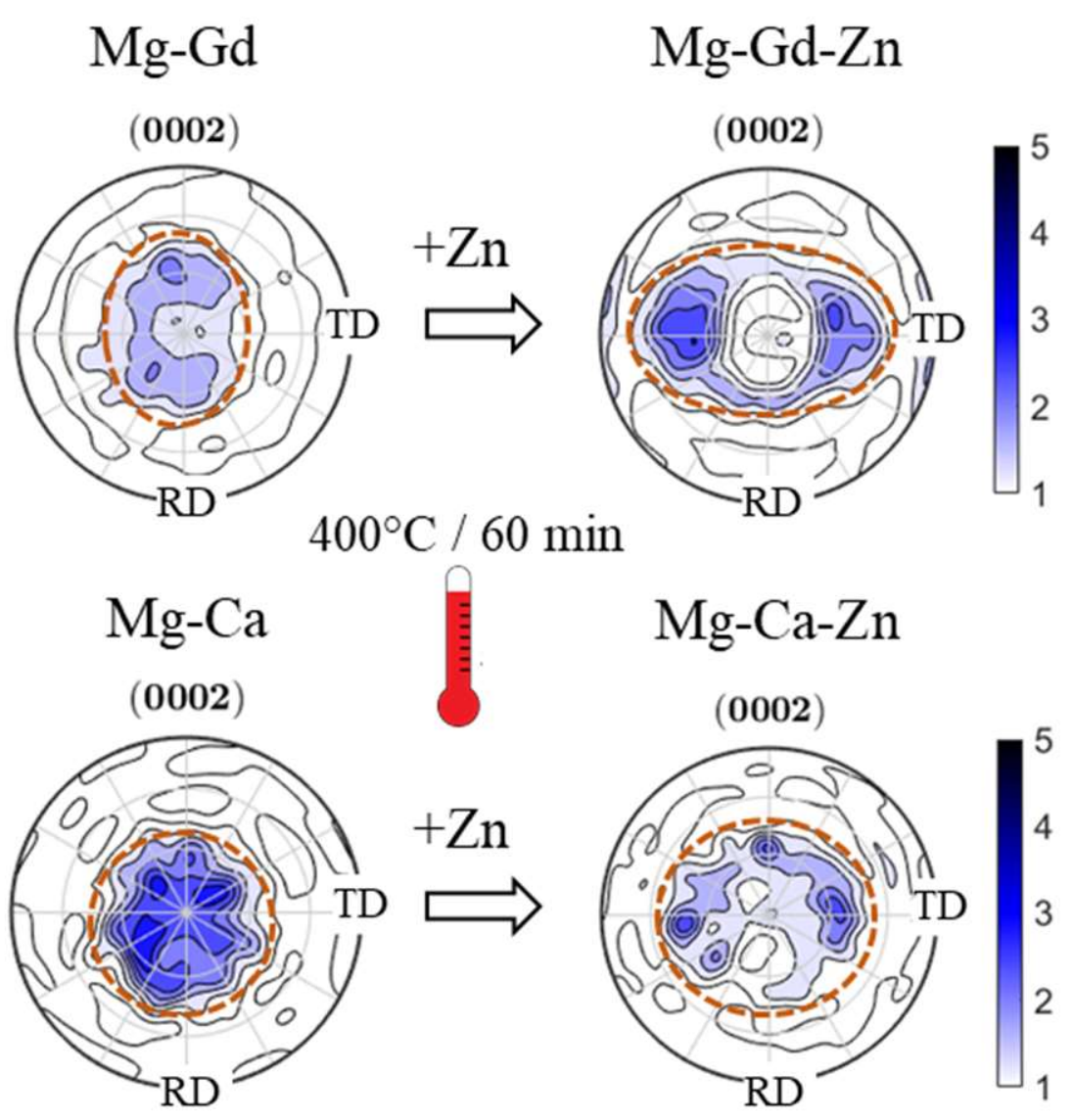

**Figure 1:** Comparison of the texture development in Mg-Gd-(Zn) and Mg-Ca-(Zn) alloys upon annealing at 400°C for 60 min. The color coding for the texture intensities is displayed in terms of multiples of a random distribution. The outlined areas in the (0002) pole figures denote the texture spread with respect to the RD and TD directions.

### *3.2. Solute segregation*

The spatial distribution of solute atoms at general, high-angle grain boundaries was investigated by means of 3D APT measurements. An overview of the grain boundary angles and axes is given in Table 2.

Table 2. Angles and Axes of the general, high angle grain boundaries in Mg-Ca-(Zn) and Mg-Gd-(Zn) investigated by 3D APT.

| Alloys | Angle (°) | Axis (hkil) |
|---|---|---|
| Mg-Gd | 72.3 | $\bar{2}\ 3\ \bar{1}\ 1$ |
| Mg-Ca | 66.6 | $\bar{1}\ 2\ \bar{1}\ 1$ |
| Mg-Gd-Zn | 79 | $\bar{5}\ 6\ \bar{1}\ 0$ |
| Mg-Ca-Zn | 66.4 | $\bar{2}\ 3\ \bar{1}\ 1$ |

The reconstructed tips and solute concentration profiles across the measured GB segments are displayed in Fig. 2 and 3 for the Mg-Gd-(Zn) and Mg-Ca-(Zn) alloys, respectively. In the case of the Gd-containing alloys, the boundary concentration of Gd in the binary alloy was relatively low (Fig. 2c) in comparison to the ternary alloy upon the introduction of Zn (Fig. 2d) (0.5 at.% vs. 1.5 at.%). Additionally, a significant amount of Zn appears to co-segregate at the GB (2.0 at.%) (Fig. 2d). In the Ca-containing alloys, the GB segregation behavior of Ca was similar in both alloys (2.3 at.% in Mg-Ca vs. 1.9 at.% in Mg-Ca-Zn). Accordingly, the addition of Zn did not seem to enhance the segregation behavior of Ca. The boundary concentration of Zn was lower (0.8 at.%), compared to the Mg-Gd-Zn alloy.

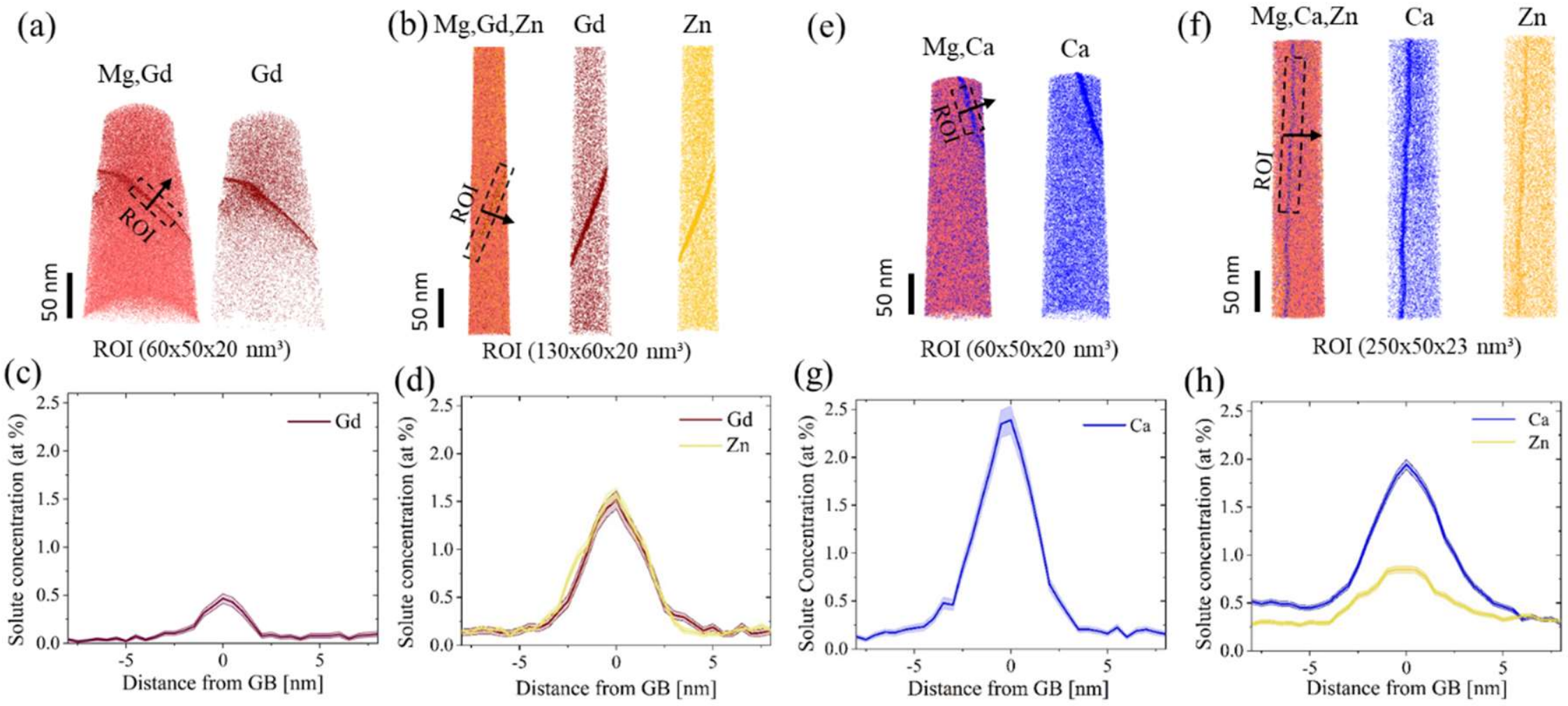

**Figure 2:** APT results of the measured tips containing general, high-angle boundaries. (a,b,e,f) 3D elemental distributions of Mg and the solute atoms in the reconstructed tips of the Mg-Gd-(Zn) and Mg-Ca-(Zn) alloys. (c,d,g,h) 1D concentration profiles across the GB. The measurement position and direction are outlined by rectangular regions of interest (ROI) and a marked direction, perpendicular to the GB plane. The GB misorientation angles/axes were the following: $72°\langle\bar{2}3\bar{1}1\rangle$ (a), $79°\langle\bar{5}6\bar{1}0\rangle$ (b), $66°\langle\bar{1}2\bar{1}1\rangle$ (c), $66°\langle\bar{2}3\bar{1}1\rangle$ (d).

*3.3. Solute clustering*

Whether solute atoms are arranged randomly in the solid solution or exhibit specific spatial patterns due to clustering, this aspect can significantly influence the material properties. In order to quantify the measured solute distributions, we employed the clustering ratio $R_{A-B}$ parameter to detect possible solute clusters in the atom probe point cloud data, and examine how the clustering tendency of RE and Ca atoms is influenced by the addition of Zn in the ternary alloys. Fig. 3 shows the $R_{A-B}$ values in the binary and ternary alloys as a function of the cut-off radius for the entire APT tip (solid lines), GB region (short-dashed lines), and the bulk (long-dashed lines). As shown, the $R_{A-B}$ values decrease with an increasing cut-off radius, eventually approaching 1 in the bulk regions. This trend indicates a homogeneous distribution when a sufficiently large space of the measured tips is sampled. By comparison, the $R_{A-B}$ values for the GB region are consistently higher than those for the bulk region, suggesting a more inhomogeneous distribution of solute atoms in the vicinity of grain boundaries in general. Fig. 3a shows a comparison of the Gd-Gd vs. Ca-Ca cluster distribution in the Mg-X system. Note that for the GB region, the $R_{\mathrm{Gd-G}}$ value is higher than the $R_{\mathrm{Ca-C}}$ value within a distance $r < 4.5$ Å corresponding to second nearest-neighbor (2NN) sites (lattice constant $a$ = 3.2 Å). This suggests a less uniform distribution of Gd compared to Ca solutes in the GB for short-range order. In the bulk region, the $R_{\mathrm{Gd-Gd}}$ value is higher than the $R_{\mathrm{Ca-Ca}}$ value for all distances. Interestingly, with the addition of Zn (Fig. 3b), the clustering tendency of Ca-Ca solutes in the GB region decreases dramatically, as compared to the binary system. Accordingly, the $R_{\mathrm{Gd-G}}$ value for the GB region is significantly higher than the $R_{\mathrm{Ca-Ca}}$ value for all the distances considered. As for clustering of Zn-Zn pairs in the ternary alloys, $R_{\mathrm{Zn-Zn}}$ in the Mg-Gd-Zn alloy was higher than in the Mg-Ca-Zn alloy (Fig. 3c). For the latter, the inhomogeneity of Zn solute distribution in the GB region is comparable to that in the bulk, whereas in the Mg-Gd-Zn alloy, the inclination of Zn-Zn pairs to cluster is more pronounced in the GB region than in the bulk. In Fig. 3d, the plotted $R_{\mathrm{Zn-X}}$ ratio characterizes the tendency for X solute species to cluster around Zn sites, indicating the attraction of Zn solutes to X solutes. As evident, the $R_{\mathrm{Zn-Gd}}$ ratio notably exceeds the $R_{\mathrm{Zn-Ca}}$ ratio for all the distances considered, particularly in the GB region (cf. Fig. 3d). This suggests that clustering of Zn-Gd pairs is more favorable than the clustering of Zn and Ca.

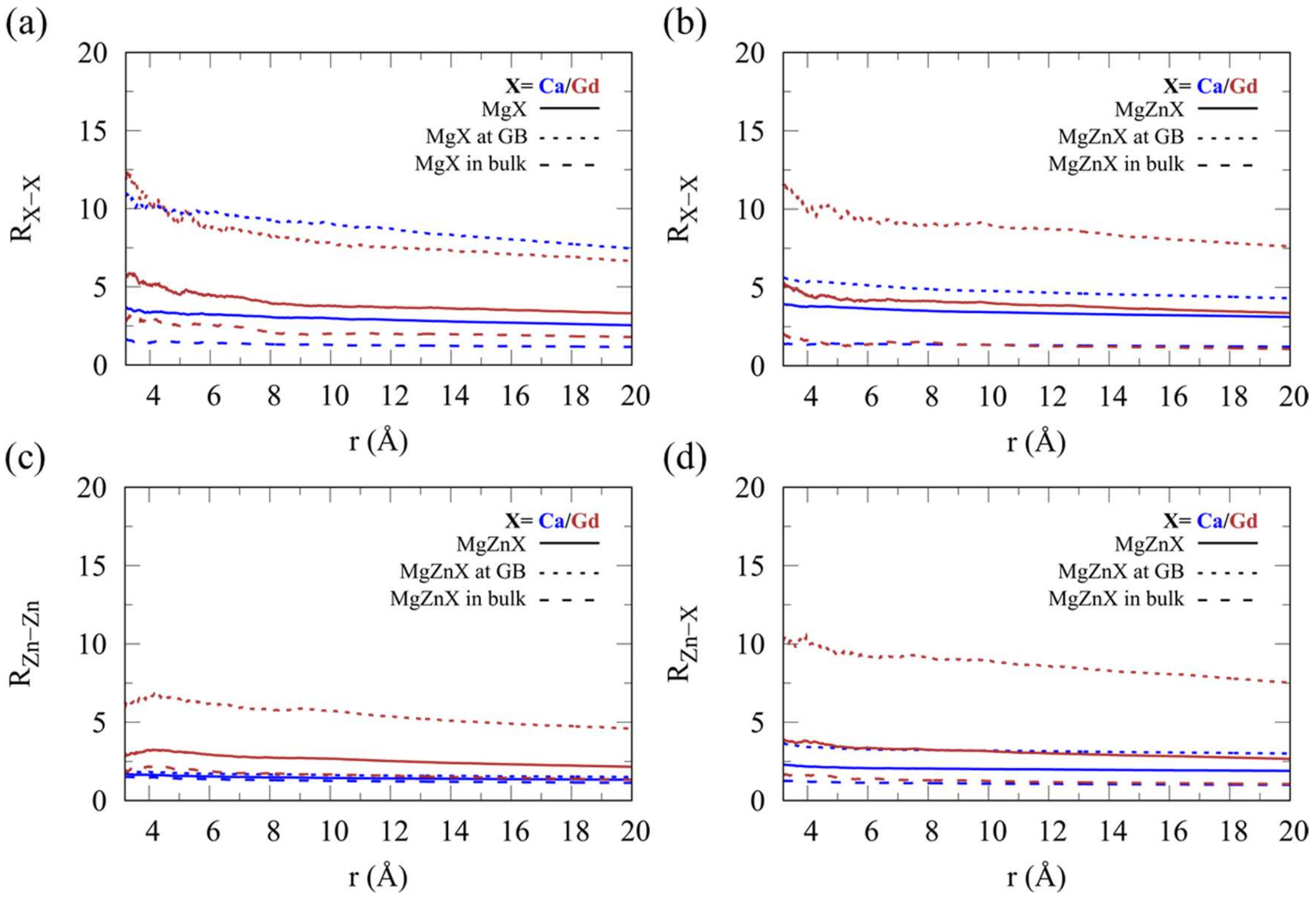

**Figure 3:** Clustering analysis of the solute distribution in the reconstructed APT samples as a function of a cutoff distance *r*. (a) Clustering of X-X atom pairs in binary Mg-X alloys. (b) Clustering of X-X pairs, (c) Zn-Zn pairs, and (d) Zn-X pairs in ternary Mg-X-Zn alloys. X = Gd or Ca.

### *3.4. Solute binding*

The binding energies of vacancy-X and Zn-X pairs in the Mg matrix were computed using DFT calculations. The vacancy-X binding energy served as an estimate for the binding of solute X to a GB site with excess free volume compared to the matrix. Meanwhile, the interaction of Zn and X solutes in the ternary alloys was evaluated based on the Zn-X binding energy within the Mg matrix. Fig. 4a shows examples of the orthorhombic simulation cells containing X solute (e. g. Ca) and Zn, as well as a vacancy and X solute (e. g. Gd). When the center feature, i.e. a vacancy or a solute, is placed at the origin of the supercell, there are two possible positions for a first nearest neighbor site. One is on the basal plane with a pair distance of $a_0$, and the second is on the prismatic plane with a pair distance of $0.577a_0 + 0.5c_0$. Both possibilities were considered when computing the binding energies, resulting in two different energies for each case.

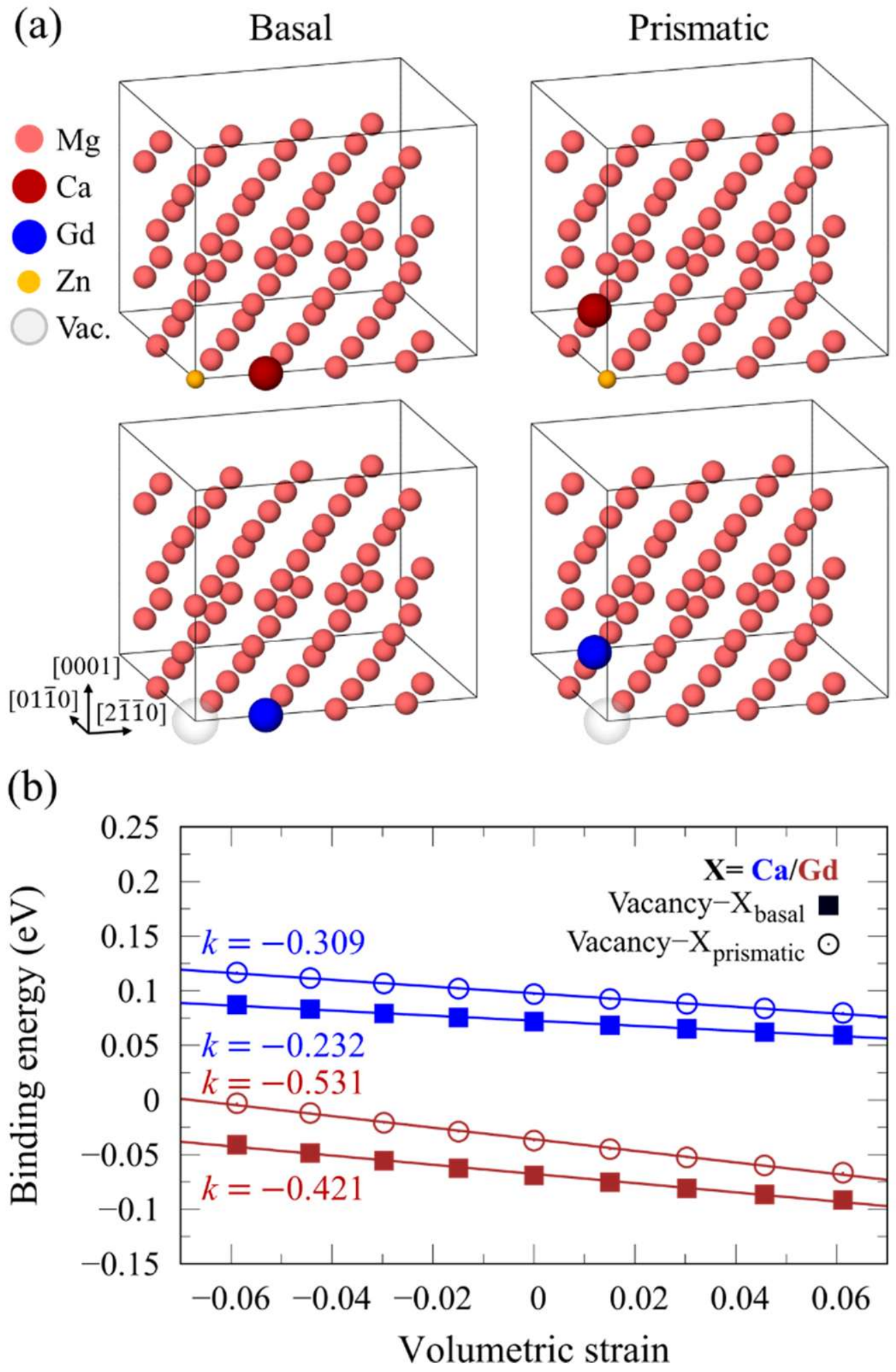

**Figure 4:** DFT calculations *of the* solute-solute and solute-vacancy binding energies in Mg. (a) Orthorhombic simulation cells containing Zn-X and vacancy-X pairs, where *either* Zn or *a* vacancy is *placed* at the origin of the supercell and *the* solute atom X is placed at the first nearest neighbor on the basal or the prismatic plane. *X=Gd or Ca.* (b) Volumetric strain-dependent vacancy-X binding energies in 5 × 3 × 3 unit cells (180 atoms). The results were fitted to linear regression with *k* as the gradient.

Table 2 summarizes the resulting binding energies for vacancy-X and Zn-X pairs in the binary and ternary alloys. For the binary Mg-X alloys, it was observed that Gd binding to a vacancy was notably less favorable compared to Ca binding to a vacancy. This is evident from the negative value of the vacancy-Gd binding energy, indicating that this binding is energetically unfavorable. In the ternary Mg-X-Zn alloys, the binding energies of Gd and Zn were larger than those of Ca and Zn. This highlights a stronger attraction between Gd-Zn pairs compared to Ca-Zn pairs. Notably, the energy associated with binding to a neighbor on the basal plane is consistently less favorable for all cases compared to the prismatic plane.

**Table 1:** Ab initio Zn-X and vacancy-X binding energies in 5×3×3 unit cells (180 atoms).

| X | $E_{\mathbf{bind}}^{\mathbf{vac.\text{-}X}}$ (eV) | $E_{\mathbf{bind}}^{\mathbf{Zn\text{-}X}}$ (eV) |
|---|---|---|
| $Ca_{basal}$ | 0.072 | 0.070 |
| $Ca_{pris.}$ | 0.097 | 0.080 |
| $Gd_{basal}$ | -0.069 | 0.083 |
| $Gd_{pris.}$ | -0.037 | 0.099 |

The volumetric strain-dependent binding energies for vacancy-X pairs were also computed, as illustrated in Fig. 4b. The vacancy-X pairs are energetically more favorable under compressive rather than tensile strain, as evidenced by the linear decrease in binding energies with strain. Interestingly, the binding energies of the vacancy-Ca pairs remain positive even at the maximum tensile strain of 6%, considered in this work. In contrast, the binding energies of vacancy-Gd pairs appear to be more sensitive to the volumetric strain, with the gradient of the fitted linear function being 1.7 to 1.8 times higher than that of the vacancy-Ca pairs. It is worth noting that the binding energy is anticipated to become positive for vacancy-Gd pairs on the prismatic plane when the compressive strain exceeds 7%.

## 4 Discussion

The findings presented in this study highlight important distinctions in the clustering and chemical binding behavior of Ca and Gd in the presence of vacancy-rich defects, like grain boundaries. Synergistic ternary effects arising from the co-addition of transition elements, such as Zn, were also observed to modify the clustering and binding tendencies of solutes, as compared to the binary systems. DFT calculation results revealed that the binding between vacancy sites and Ca solute is energetically more favorable in comparison to Gd. This explains why, in the Mg-Ca alloy, there were high levels of Ca segregation at the GB, and a less pronounced texture spread in RD compared to the Mg-Gd alloy counterpart. Moreover, it was found that the binding energies between vacancies and Gd solute exhibit high sensitivity to volumetric strain, which would cause an inhomogeneous segregation pattern, with a preference for GB sites under compressive volumetric strain condition. This observation is consistent with the clustering analysis of the APT point cloud data, which illustrate a less uniform Gd solute distribution within a 2NN distance (cf. Fig. 3a). In this context, the presence of an inhomogeneous solute distribution within GBs is likely to influence their mobilities, subsequently impacting the development of texture.

In the ternary alloys, the formation of Gd-Zn pairs proved to be more favorable compared to Ca-Zn pairs. This preference led to a more pronounced synergistic effect of Zn with Gd, consequently enhancing their co-segregation behavior. This effectively explains the similarly high concentration of both Gd and Zn at the GB (Fig. 2d), as well as their pronounced clustering in the GB region (Fig. 3d). Conversely, Ca-Zn pairs exhibited a lower binding energy, which does not seem to strongly influence the individual segregation of Ca to the GB (Fig. 2g) but rather affect the segregation of Zn (Fig. 2h). This observation strongly suggests that the synergistic effect of combined alloying additions depends on their binding behavior. The more attractive the binding, the more effective their synergy in GB co-segregation and texture modification. For additional insights into the role of combined solute addition in promoting the segregation levels, the segregation energy $\Delta G_{seg}$ in all four alloys was calculated after the Langmuir-McLean adsorption model represented by Eq. 4.1. $X_{GB}$ represents the peak solute concentration at the grain boundary, and $X_M$ the solute concentration in the matrix [30].

$$\frac{X_{GB}}{1-X_{GB}} = \left(\frac{X_M}{1-X_M}\right) \times \exp\left(-\frac{\Delta G_{seg}}{RT}\right) \qquad \text{(Eq. 4.1)}$$

Fig. 5a and b depict a correlation between the peak concentration in the GB and the segregation energy $\Delta G_{seg}$ for both binary and ternary alloys. The segregation energies corroborate the solute segregation behavior indicated by the peak concentrations. For instance, in the case of Mg-Gd-Zn , the negative segregation energies of both Gd and Zn are approximately twice as high as those of Ca and Zn in the Mg-Ca-Zn alloy . Conversely, in the binary alloys, the negative segregation energy of Ca is higher than that of Gd (Fig. 5b).

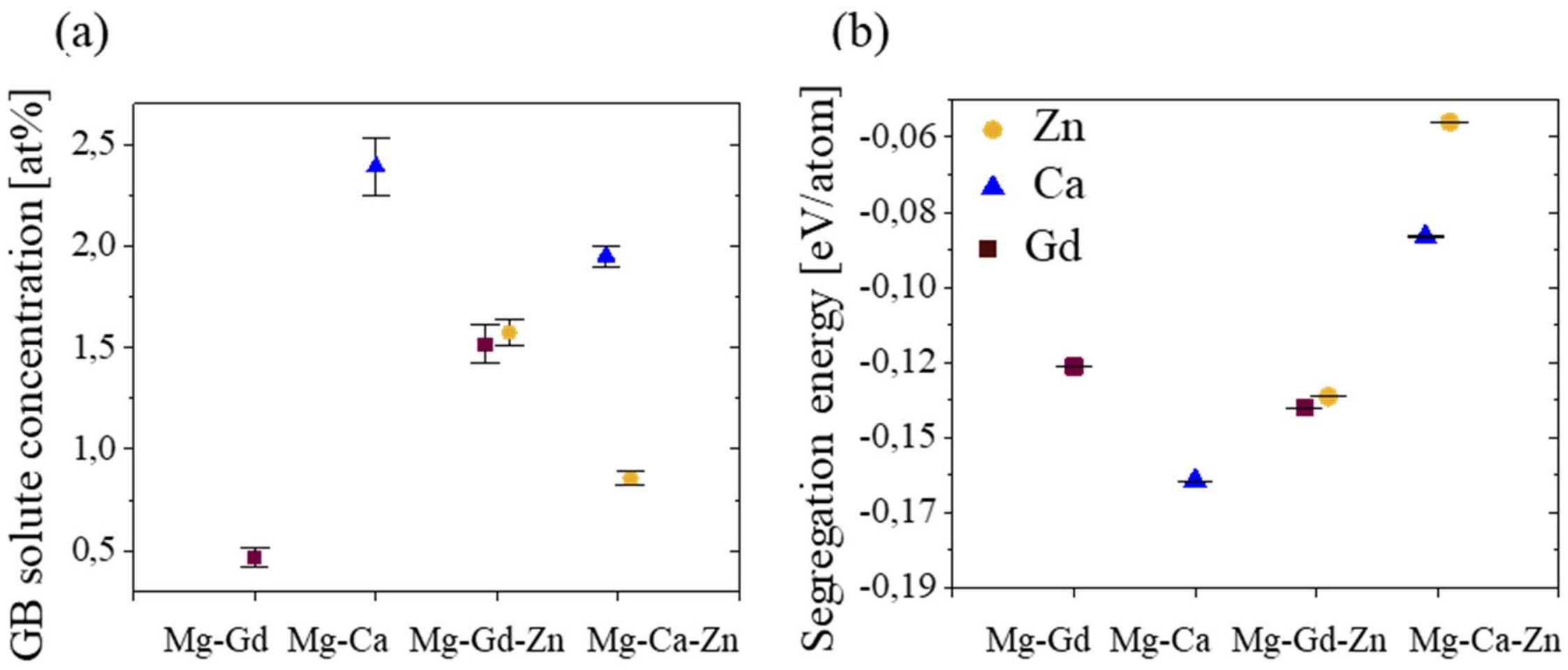

**Figure 5:** (a) Peak concentration of solute at the GB in Mg-Gd , Mg-Ca , Mg-Gd-Zn and Mg-Ca-Zn . (b) The corresponding GB segregation energy.

The observed characteristics in solute-vacancy and solute-solute interactions with respect to binding and segregation are likely to underlie the distinct TD-spread tendency of the annealing texture observed in the Mg-RE-Zn alloy. Notably, the “boost” in the segregation behavior of Gd upon the addition of Zn, causes a pronounced shift of the orientation of maximum intensity in the (0002) pole figure from the RD towards the TD (Fig. 1). This aligns with the literature, which has previously proposed a connection between the spread of texture poles and the segregation of solutes [8, 31-33]. However, the current data uniquely reveals that the association between the formation of a TD-split texture and the magnitude of segregation benefits largely from the strong Zn/Gd solute synergy present in the Mg-Gd-Zn alloy. In this regard, it is important to point out that despite a relatively high solute concentration of Ca at the GB in the Mg-Ca-Zn alloy (2 at.%), the resulting annealing texture does not exhibit a significant TD-spread feature akin to that observed in the Mg-Gd-Zn alloy. This is likely due to the weaker attraction between Zn and Ca atoms, resulting in a less pronounced segregation of Zn to the grain boundaries.

The findings presented in this paper provide substantial evidence supporting the notion that the preferential development of specific texture components in Mg alloys is intricately connected to solute segregation through heterogeneous solute-boundary interactions. The reliance of binding energies on volumetric strains implies that solute segregation at grain boundaries is influenced by the local grain boundary structure. Consequently, a non-uniform GB solute distribution leads to varying grain boundary mobilities among different boundaries, facilitating the selective growth of specific texture components. These results align with previous literature that has also proposed anisotropic grain boundary segregation in diverse material systems, exemplary Mg-Mn-Nd, Au-Pt and Fe-B-C [3, 16, 34].

In the design of wrought magnesium alloys, it is crucial to recognize that achieving texture modification via the incorporation of RE/Ca and Zn alone may not be effective unless there is a sufficient driving force for co-segregation. In this context, the presence of a strong chemical interaction between RE/Ca and Zn emerges as a pivotal factor. Therefore, the segregation behavior and the redistribution of solutes are not solely determined by the minimization of elastic strain energy but are also heavily influenced by the strength of the chemical binding between these elements. The latter aspect is in excellent agreement with a very recent study that employed atomistic simulations to illustrate that solute segregation or depletion in Mg-Al alloys is induced by vacancy clusters [35].

## 5 Conclusions

The current study has contributed to a clearer understanding of the impact of chemical binding and solute distribution on the segregation behavior at grain boundaries, and the subsequent alteration of the annealing texture in dilute Mg-X-Zn alloys (X= RE or Ca). The following conclusions can be drawn:

1. This study reveals key differences in how Ca and Gd behave in the presence of vacancy-rich defects like grain boundaries. DFT calculations showed that Ca binds more favorably to vacancy sites than Gd, explaining the higher Ca segregation at general grain boundaries and the resulting RD-TD spread texture.
2. The addition of a transition element (Zn) had significant synergistic effects on solute clustering and binding. Gd-Zn pairs formed more favorably than Ca-Zn pairs, leading to a stronger synergistic effect between Zn and Gd. This in turn explains their equally high concentration at the grain boundary, and the resulting TD spread texture. In contrast, Ca-Zn pairs exhibited weaker binding, showing no influence on Ca segregation to the grain boundary but weakend Zn segregation.

3. The DFT predictions and experiments offer substantial evidence that the selective formation of specific texture components in Mg alloys is closely tied to heterogeneous solute-boundary interactions. The fact that the binding energy can be sensitive to the volumetric strain suggests that solute segregation at grain boundaries is influenced by the local grain boundary structure. Consequently, an uneven distribution of solutes along grain boundaries results in varying grain boundary mobilities, enabling the preferential growth of specific texture components.
4. Solute behavior regarding clustering and segregation is shaped not only by the large atomic size of RE and Ca, and the resulting elastic strain energy minimization, but also by the considerable impact of the chemical binding strength, either with vacancies or with co-added Zn in the ternary alloys. In terms of alloy design, this highlights a crucial aspect that the efficacy of alloying elements depends on the binding behavior of co-added solutes. In other words, stronger binding enhances the synergy for grain boundary co-segregation and texture modification.

**Acknowledgements**

F.M., and T.A.S. are grateful for the financial support from the German Research Foundation (DFG) (Grant Nr. AL1343/7-1). Z.X. and T.A.S. acknowledge the financial support by the DFG (Grant Nr. 505716422). Z.X. and S.K.K. acknowledge the financial support by the DFG through the SFB1394 Structural and Chemical Atomic Complexity – From Defect Phase Diagrams to Material Properties, project ID 409476157. J.G. acknowledges funding from the French National Research Agency (ANR), Grant ANR-21-CE08-0001 (ATOUUM) and ANR-22-CE92-0058-01 (SILA). Simulations were performed with computing resources granted by RWTH Aachen University under project (p0020267).

## Literature

1. Bhattacharyya, J.J., S.R. Agnew, and G. Muralidharan, *Texture enhancement during grain growth of magnesium alloy AZ31B.* Acta Materialia, 2015. **86**: p. 80-94.
2. Pei, R., et al., *Solute drag-controlled grain growth in magnesium investigated by quasi in-situ orientation mapping and level-set simulations.* Journal of Magnesium and Alloys, 2023. **11**: p. 2312-2325.
3. Zhou, X., et al., *Atomic motifs govern the decoration of grain boundaries by interstitial solutes.* Nat Commun, 2023. **14**(1): p. 3535.
4. Barrett, C.D., H. El Kadiri, and R. Moser, *Generalized interfacial fault energies.* International Journal of Solids and Structures, 2017. **110-111**: p. 106-112.
5. Basu, I., T. Al Samman, and G. Gottstein, *Recrystallization and Grain Growth Related Texture and Microstructure Evolution in Two Rolled Magnesium Rare-Earth Alloys.* Materials Science Forum, 2013. **765**: p. 527-531.
6. Hadorn, J.P., et al., *Role of Solute in the Texture Modification During Hot Deformation of Mg-Rare Earth Alloys.* Metallurgical and Materials Transactions A, 2011. **43**(4): p. 1347-1362.
7. Mouhib, F.-Z., et al., *Texture Selection Mechanisms during Recrystallization and Grain Growth of a Magnesium-Erbium-Zinc Alloy.* Metals, 2021. **11**(1).
8. Nie, J.F., et al., *Solute segregation and precipitation in a creep-resistant Mg–Gd–Zn alloy.* Acta Materialia, 2008. **56**(20): p. 6061-6076.
9. Imandoust, A., et al., *A review on the effect of rare-earth elements on texture evolution during processing of magnesium alloys.* Journal of Materials Science, 2016. **52**(1): p. 1-29.
10. Shao, X.H., et al., *Atomic-scale segregations at the deformation-induced symmetrical boundary in an Mg-Zn-Y alloy.* Acta Materialia, 2016. **118**: p. 177-186.
11. Nie, J.F., et al., *Periodic segregation of solute atoms in fully coherent twin boundaries.* Science, 2013. **340**(6135): p. 957-60.
12. He, C., et al., *Unusual solute segregation phenomenon in coherent twin boundaries.* Nat Commun, 2021. **12**(1): p. 722.
13. Basu, I. and T. Al-Samman, *Triggering rare earth texture modification in magnesium alloys by addition of zinc and zirconium.* Acta Materialia, 2014. **67**: p. 116-133.
14. Guan, D., et al., *Exploring the mechanism of "Rare Earth" texture evolution in a lean Mg–Zn–Ca alloy.* Scientific Reports, 2019. **9**: p. 7152.
15. Pei, R., et al., *Grain boundary co-segregation in magnesium alloys with multiple substitutional elements.* Acta Materialia, 2021. **208**: p. 116749.
16. Pei, R., et al., *Atomistic origin of the anisotropic grain boundary segregation in a Mg-Mn-Nd alloy.* Archived on arXiv, 2022.
17. Mendelev, M.I. and D.J. Srolovitz, *Co-Segregation Effects on Boundary Migration.* Interface Science, 2002. **10**: p. 191-199.
18. Mahjoub, R. and N. Stanford, *The electronic origins of the "rare earth" texture effect in magnesium alloys.* Sci Rep, 2021. **11**(1): p. 14159.
19. Barrett, C.D., A. Imandoust, and H. El Kadiri, *The effect of rare earth element segregation on grain boundary energy and mobility in magnesium and ensuing texture weakening.* Scripta Materialia, 2018. **146**: p. 46-50.
20. Mouhib, F., et al., *Synergistic effects of solutes on active deformation modes, grain boundary segregation and texture evolution in Mg-Gd-Zn alloys.* Materials Science and Engineering: A, 2022. **847**: p. 143348.
21. Du, Y.Z., et al., *Effect of microalloying with Ca on the microstructure and mechanical properties of Mg-6 mass%Zn alloys.* Materials & Design, 2016. **98**: p. 285-293.

22. Zeng, Z.R., et al., *Effects of dilute additions of Zn and Ca on ductility of magnesium alloy sheet.* Materials Science and Engineering: A, 2016. **674**: p. 459-471.
23. Hielscher, R. and H. Schaeben, *A novel pole figure inversion method: specification of theMTEXalgorithm.* Journal of Applied Crystallography, 2008. **41**(6): p. 1024-1037.
24. Tilocca, A., *Structure and dynamics of bioactive phosphosilicate glasses and melts from ab initio molecular dynamics simulations.* Physical Review B, 2007. **76**(22): p. 224202.
25. Giannozzi, P., et al., *QUANTUM ESPRESSO: a modular and open-source software project for quantum simulations of materials.* J Phys Condens Matter, 2009. **21**(39): p. 395502.
26. Kresse, G. and D. Joubert, *From ultrasoft pseudopotentials to the projector augmented-wave method.* Physical Review B, 1999. **59**(3): p. 1758-1775.
27. Blöchl, P.E., *Projector augmented-wave method.* Physical Review B, 1994. **50**(24): p. 17953-17979.
28. Perdew, J.P., K. Burke, and M. Ernzerhof, *Generalized Gradient Approximation Made Simple.* Physical Review Letters, 1996. **77**(18): p. 3865-3868.
29. Fletcher, R., *Practical methods of optimization*. 2000: John Wiley & Sons.
30. McLean, D. and A. Maradudin, *Grain Boundaries in Metals.* Physics Today, 1958. **11**(7): p. 35-36.
31. Mouhib, F.-Z., et al., *Texture Selection Mechanisms during Recrystallization and Grain Growth of a Magnesium-Erbium-Zinc Alloy.* Metals, 2021. **11**(1): p. 171.
32. Jiang, M.G., et al., *Correlation between dynamic recrystallization and formation of rare earth texture in a Mg-Zn-Gd magnesium alloy during extrusion.* Sci Rep, 2018. **8**(1): p. 16800.
33. Jiang, M.G., et al., *Quasi-in-situ observing the rare earth texture evolution in an extruded Mg-Zn-Gd alloy with bimodal microstructure.* Journal of Magnesium and Alloys, 2020.
34. Barr, C.M., et al., *The role of grain boundary character in solute segregation and thermal stability of nanocrystalline Pt-Au.* Nanoscale, 2021. **13**(6): p. 3552-3563.
35. Yi, P., et al., *The interplay between solute atoms and vacancy clusters in magnesium alloys.* Acta Materialia, 2023. **249**: p. 118805.
36. Tzanetakis, P., J. Hillairet, and G. Revel, *The formation energy of vacancies in aluminium and magnesium.* physica status solidi (b), 1976. **75**(2): p. 433-439.
37. Mairy, C., J. Hillairet, and D. Schumacher, *Energie de formation et concentration d'équilibre des lacunes dans le magnésium.* Acta metallurgica, 1967. **15**(7): p. 1258-1261.
38. Shin, D. and C. Wolverton, *First-principles study of solute–vacancy binding in magnesium.* Acta Materialia, 2010. **58**(2): p. 531-540.

# Supplementary Materials

**Table S1:** Ab initio vacancy formation energies in 4×2×2 unit cells (64 atoms) and 5×3×3 unit cells (180 atoms).

| **Method** | **Distance (Å)** | $E_{vac.}$ **(eV)** |
|---|---|---|
| Cell 4×2×2, K-point 6×6×6 | $3.26a_0$ | 0.793 |
| Cell 5×3×3, K-point 4×4×4 | $4.89a_0$ | 0.807 |
| Literature, experiments | - | 0.79±0.03 [36]<br>0.81±0.02 [37] |
| Literature, DFT | - | 0.74 [38]<br>0.83 [38] |

**Table S2:** Ab initio Zn-X and vacancy-X binding energies in 4×2×2 unit cells (64 atoms) and 5×3×3 unit cells (180 atoms).

| **X** | **Cell** | **K-point** | **Distance (Å)** | $E_{bind}^{vac.\text{-}X}$ **(eV)** | $E_{bind}^{Zn\text{-}X}$ **(eV)** |
|---|---|---|---|---|---|
| $Ca_{basal}$ | 4×2×2 | 6×6×6 | $3a_0$ | 0.058 | 0.078 |
| $Ca_{basal}$ | 5×3×3 | 4×4×4 | $4a_0$ | 0.072 | 0.070 |
| $Ca_{pris}$ | 4×2×2 | 6×6×6 | $2.51a_0$ | 0.081 | 0.089 |
| $Ca_{pris}$ | 5×3×3 | 4×4×4 | $4.11a_0$ | 0.097 | 0.080 |
| $Gd_{basal}$ | 4×2×2 | 6×6×6 | $3a_0$ | -0.064 | 0.084 |
| $Gd_{basal}$ | 5×3×3 | 4×4×4 | $4a_0$ | -0.069 | 0.083 |
| $Gd_{pris.}$ | 4×2×2 | 6×6×6 | $2.51a_0$ | -0.051 | 0.103 |
| $Gd_{pris.}$ | 5×3×3 | 4×4×4 | $4.11a_0$ | -0.037 | 0.099 |